\documentclass[preprint,showpacs]{revtex4}
\usepackage{epsfig}
\usepackage{amsmath}
\usepackage{amssymb}

\begin{document}

\preprint{BIHEP-TH-2004-5}

\title{Branching ratio and CP asymmetry  of $B_{s}  \to \pi^{+}\pi^{-}$ \\ decays
in the perturbative QCD approach}

\author{Ying Li$^{a,b,c}$ Cai-Dian L\"u$^{a,b}$ ZhenJun Xiao$^{a,c}$  Xian-Qiao Yu$^{b}$}

\affiliation{{\it \small $a$ CCAST (World Laboratory), P.O. Box 8730,
Beijing 100080, China}\\
{\it \small $b$   Institute of High Energy Physics,
P.O.Box 918(4), Beijing 100039, China}\\
{\it \small $c$ Physics Department, NanJing Normal University,
JiangSu 210097, China}}

\begin{abstract}
In this paper, we calculate the decay rate and CP asymmetry of the
$B_s \to \pi^+\pi^-$ decay in perturbative QCD approach with
Sudakov resummation. Since none of the quarks in final states is
the same as those of the initial $B_s$ meson, this decay can occur
only via annihilation diagrams in the standard model. Besides the
current-current operators, the contributions from the QCD and
electroweak penguin operators are also taken into account. We find
that (a) the branching ratio is about $4 \times 10^{-7}$; (b) the
penguin diagrams dominate the total contribution; and (c) the
direct CP asymmetry is small in size: no more than $3\% $; but the
mixing-induced CP asymmetry can be as large as ten percent
testable in the near future LHC-b experiments.
\end{abstract}

 \maketitle

\newpage
\section{Introduction}

   In recent years,  rare $B$ decays are attracting  more and more attentions,
 since they provide a good opportunity  for testing the Standard
 Model(SM), probing $CP$ violation and searching for possible new physics beyond the SM.
 Since 1999, the data sample of the pair production and decays of $B$ mesons
 collected by BaBar and Belle Collaborations is increased rapidly.
 In the future LHC-b experiments, $B_s$ and $B_C$
 mesons can also be produced, and the rare $B$ decays with a branching ratio around $10^{-7}$ can be
 observed.  The rapid progress in current $B$ factory experiments and the bright expectation in LHC-b
 experiments induced a great  interest in  the studies of rare decays of $B$ meson.

 The rare decay $B_{s}\rightarrow  \pi^{+}\pi^{-}$ can occur only via
 annihilation diagrams in SM because none of quarks in final states is
 the same as those of the initial $B_{s}$ meson. The usual method to treat
 non-leptonic decays of B meson is Factorization Approach(FA) \cite{1},
 which has achieved great success in explaining many decay
 branching ratios \cite{2,3,4}. However, this method failed in describing $B_{s}\rightarrow
 \pi^{+}\pi^{-}$ decay, because we need the $\pi\rightarrow\pi$
 form factor at very large momentum transfer ${\cal O} (M_{B})$.
 So far, little is known about the form factor at such a large momentum transfer
in FA. In the QCD factorization approach \cite{5}, one cannot perform
a real calculation of the annihilation diagrams, but estimating  the annihilation amplitude
by introducing a phenomenological parameter. In this paper, we calculate the
branching ratio and $CP$ asymmetries of $B_{s}\rightarrow\pi^{+}\pi^{-}$ decay by
employing the perturbative QCD approach(PQCD) \cite{6}. This method  has been developed for the studies
of the B meson decays \cite{7} and successfully applied to calculate the annihilation
diagrams \cite{8,9}. When the final states are light mesons
such as pions, the perturbative QCD approach(PQCD) can be safely used
because of asymptotic freedom of QCD \cite{10}.

In the next section, we give our theoretical formulas for the
decay $B_{s}\rightarrow\pi^{+}\pi^{-}$ in PQCD framework. In
section 3, we give the numerical results of the branching ratio of
$B_{s}\rightarrow\pi^{+}\pi^{-}$ and discuss CP asymmetry of the
decay.

\section{Perturbative calculations}\label{sc:fm}

The related effective Hamiltonian for the process
$B_{s}\rightarrow\pi^{+}\pi^{-}$ is given by \cite{9,11}
\begin{equation}
 H_\mathrm{eff} = \frac{G_F}{\sqrt{2}}\{ V_{ub}^*V_{us} \left[
C_1(\mu) O_1(\mu) + C_2(\mu) O_2(\mu)
\right]-V_{tb}^*V_{ts}\sum_{i=3}^{10}C_i(\mu)
O_i(\mu)\},\label{hami}
\end{equation}
 where $C_{i}(\mu)(i=1,\cdots,10)$ are Wilson coefficients at the
  renormalization scale $\mu$ and the operators $O_{i}(i=1,\cdots,10)$ are
\begin{gather}
  O_1 = (\bar{s}_iu_j)_{V-A}(\bar{u}_jb_i)_{V-A},\hspace*{6mm}\nonumber\\
  O_2 = (\bar{s}_iu_i)_{V-A} (\bar{u}_jb_j)_{V-A},\hspace*{6mm}\nonumber\\
  O_3 = (\bar{s}_ib_i)_{V-A}\sum_{q} (\bar{q}_jq_j)_{V-A},\nonumber\\
  O_4 = (\bar{s}_ib_j)_{V-A}\sum_{q} (\bar{q}_jq_i)_{V-A},\nonumber\\
  O_5 = (\bar{s}_ib_i)_{V-A}\sum_{q} (\bar{q}_jq_j)_{V+A},\nonumber\\
  O_6 = (\bar{s}_ib_j)_{V-A} \sum_{q} (\bar{q}_jq_i)_{V+A},\nonumber\\
  \hspace{6mm}O_7 = \frac{3}{2}(\bar{s}_ib_i)_{V-A} \sum_{q} e_q(\bar{q}_jq_j)_{V+A},\nonumber\\
  \hspace{6mm}O_8 = \frac{3}{2}(\bar{s}_ib_j)_{V-A}\sum_{q} e_q (\bar{q}_jq_i)_{V+A},\nonumber\\
  \hspace{6mm}O_9 = \frac{3}{2}(\bar{s}_ib_i)_{V-A}\sum_{q} e_q(\bar{q}_jq_j)_{V-A},\nonumber\\
 \hspace{6mm} O_{10} = \frac{3}{2}(\bar{s}_ib_j)_{V-A}\sum_{q} e_q(\bar{q}_jq_i)_{V-A}.
\end{gather}
Here $i$ and $j$ are $SU(3)$ color indices, the sum over $q$ runs
over the quark field that are active at the scale $\mu=O(m_{b})$,
i.e., $q\in \{u,d,s,c,b\}$. Operators $O_{1}, O_{2}$ come from
tree level, $O_{3}, O_{4}, O_{5}, O_{6}$ are QCD-Penguins
operators and $O_{7}, O_{8}, O_{9}, O_{10}$ come from
electroweak-Penguins.

In the PQCD approach, the decay amplitude is separated into
soft($\Phi$), hard($H$), and harder($C$) dynamics characterized by
different scales. It is conceptually written as the following,
\begin{equation}
 \mbox{Amplitude}
\sim \int\!\! d^4k_1 d^4k_2 d^4k_3\ \mathrm{Tr} \bigl[ C(t)
\Phi_B(k_1) \Phi_{\pi}(k_2) \Phi_\pi(k_3) H(k_1,k_2,k_3, t)
\bigr], \label{eq:convolution1}
\end{equation}
where $k_i$'s are momenta of light quarks included in each mesons,
and $\mathrm{Tr}$ denotes the trace over Dirac and color indices.
$C(t)$ is Wilson coefficient which results from the radiative
corrections at short distance. $\Phi_{M}$ is wave function which
describes the hadronization of mesons. The wave functions should
be
universal and channel independent,  we can use $\Phi_{M}$ which is determined by other
ways. The hard part $H$ is rather process-dependent. In the following, we
start to compute the decay amplitude of
$B_{s}\rightarrow\pi^{+}\pi^{-}$.

Since we set $B_{s}$ at rest, in the light-cone coordinates, the
momentum of the $B_s$, $\pi^-$ and $\pi^+$ are written as :
\begin{equation}
       P_1 = \frac{M_B}{\sqrt{2}} (1,1,{\bf 0}_T), P_2 =
       \frac{M_B}{\sqrt{2}} (1,0,{\bf 0}_T), P_3 =
       \frac{M_B}{\sqrt{2}} (0,1,{\bf 0}_T) . \label{eq:momentun1}
\end{equation}
Denoting the light (anti-)quark momenta in $B$, $\pi^-$ and
$\pi^+$ as $k_1$, $k_2$ and $k_3$, respectively, we can choose:
\begin{equation}
k_1 = (x_1p_1^+,0,{\bf k}_{1T}), ~~~k_2 = (x_2p_2^+,0,{\bf
k}_{2T}),~~~
 k_3 = (0, x_3p_3^- ,{\bf k}_{3T}). \label{eq:momentun2}
\end{equation}

\begin{figure}[htbp]
  \begin{center}
  \epsfig{file=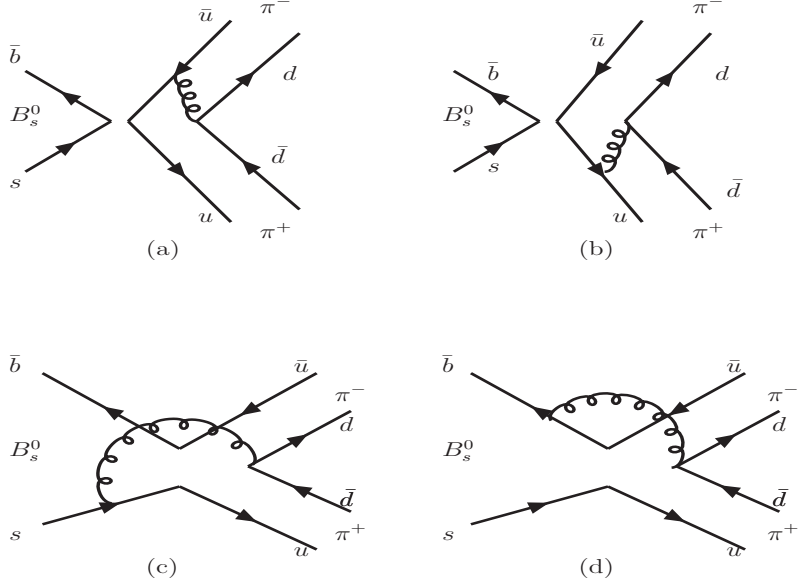,width=300pt,height=220pt}
   \end{center}
   \caption{The lowest order diagrams for $B_{s}^0 \to \pi^{+} \pi^-$ decay.}
   \label{figure:Fig1}
  \end{figure}

According to effective Hamiltonian(\ref{hami}), we draw the lowest
order diagrams of $B_{s}\rightarrow\pi^{+}\pi^{-}$ in
Fig.\ref{figure:Fig1}. For the factorizable diagrams (a) and (b),
we find their contributions cancel each other, which is a result
of exact isospin symmetry. For the non-factorizable diagrams (c)
and (d), the contribution comes from tree operator is
\begin{multline}
M_a^{T}  =  \frac{1}{\sqrt{2N_c}} 64\pi C_F M_B^2 \int_0^1\!\!\!
dx_1 dx_2 dx_3
 \int_0^\infty\!\!\!\!\! b_1 db_1\, b_2 db_2\
\phi_B(x_1,b_1) \\
\times \Bigl[     C_{2}(t_{1}) \alpha_s(t_{1})
\bigl\{-x_3\phi_{\pi}^A(x_2)\phi_{\pi}^A(x_3)-r^2(x_2+x_3)\phi_{\pi}^P(x_2)\phi_{\pi}^P(x_3)\\
+r^2(x_2-x_3)\phi_{\pi}^P(x_2)\phi_{\pi}^T(x_3)+r^2(x_2-x_3)\phi_{\pi}^P(x_3)\phi_{\pi}^T(x_2)\\
-r^2(x_2+x_3)\phi_{\pi}^T(x_2)\phi_{\pi}^T (x_3)\bigr\}
 h_a^{(1)}(x_1, x_2,x_3,b_1,b_2)
 E_{m}(t_{1}) \\
+ C_{2}(t_{2})\alpha_s(t_{2}) \bigl\{x_2\phi_{\pi}^A(x_2)\phi_{\pi}^A(x_3)+r^2(2+x_2+x_3)\phi_{\pi}^P(x_2)\phi_{\pi}^P(x_3)\\
+r^2(x_2-x_3)\phi_{\pi}^P(x_2)\phi_{\pi}^T(x_3)+r^2(x_2-x_3)\phi_{\pi}^T(x_2)\phi_{\pi}^P(x_3)\\
+r^2(-2+x_2+x_3)\phi_{\pi}^T(x_2)\phi_{\pi}^T(x_3) \bigr\}
  h_a^{(2)}(x_1, x_2,x_3,b_1,b_2)
 E_{m}(t_{2})\bigr], \label{ma}
\end{multline}
where $r=m_{0\pi}/m_B=m_{\pi}^2/[m_B(m_u+m_d)]$. $C_F=4/3$ is the
group factor of the $SU(3)_{c}$ gauge group.
     The  expressions of the meson distribution amplitudes
     $\phi_{M}$, the Sudakov factor $E_m(t_i)$ and the functions $h_a^{(1,2)}(x_1, x_2,x_3,b_1,b_2)$
    are given in the appendix.
The contribution of
penguin-diagrams $M_a^{P}$ can be obtained by replacing
$ C_{2}(t_{i})(i=1,2)$      with
 \begin{equation}
  a^{P}(t_{i})
  =2C_{4}(t_{i})+2C_{6}(t_{i})+\frac{1}{2}C_{8}(t_{i})+\frac{1}{2}C_{10}(t_{i}),
 \end{equation}
in
Eq.(\ref{ma}).
 The explicit expressions of QCD corrected Wilson coefficients $C_{2}$, $C_{4}$,
 $C_{6}$, $C_{8}$ and $C_{10}$ as a function of scale $t$ can be found in the Appendix of
 Ref.\cite{9}.

Now, the total decay amplitude for
$B_{s}\rightarrow\pi^{+}\pi^{-}$ is given by
\begin{equation}
A =
V_{ub}^*V_{us}M_{a}^{T}-V_{tb}^*V_{ts}M_{a}^{P}=V_{ub}^*V_{us}M_{a}^{T}[1+z
e^{i(\delta-\gamma)}],
\end{equation}
where $z=|V_{tb}^*V_{ts}/V_{ub}^*V_{us}||M_{a}^{P}/M_{a}^{T}|$,
$V_{ub}\simeq |V_{ub}|e^{-i\gamma}$ and $\delta$ is the relative
strong phase between tree diagrams($M_{a}^{T}$) and penguin
diagrams($M_{a}^{P}$). z and $\delta$ can be calculated from PQCD.

 The decay width is expressed as
\begin{equation}
 \Gamma(B_s^0 \to \pi^+ \pi^-) = \frac{G_F^2 M_B^3}{128\pi}
|A|^2=\frac{G_F^2
M_B^3}{128\pi}|V_{ub}^*V_{us}M_{a}^{T}|^{2}[1+z^{2}+2z\cos(\delta-\gamma)].
\label{eq:width1}
\end{equation}
Similarly, we can get the decay width for
$\bar{B}^{0}_{s}\rightarrow\pi^{+}\pi^{-}$
\begin{equation}
 \Gamma(\bar{B}_s^0 \to \pi^+ \pi^-) = \frac{G_F^2 M_B^3}{128\pi}
|\bar{A}|^2, \label{eq:width2}
\end{equation}
where
\begin{equation}
\bar{A}=V_{ub}V_{us}^{*}M_{a}^{T}-V_{tb}V_{ts}^{*}M_{a}^{P}=V_{ub}V_{us}^{*}M_{a}^{T}
[1+z e^{i(\delta+\gamma)}].
\end{equation}

\section{Numerical evaluation and summary}\label{sc:neval}

The following parameters have been used in our numerical
calculation:
\begin{gather}
\nonumber  M_{B_s} = 5.37 \mbox{ GeV},\  m_{0\pi} = 1.4 \mbox{
GeV},\  \Lambda^{f=4}_{QCD}=0.25\mbox{ GeV}, f_{B_s} = 236 \mbox{ MeV},\
\\ f_{\pi} = 130 \mbox{ MeV},\
\tau_{B_s^0}=1.46\times 10^{-12}\mbox{s},
 | V_{tb}^{*}V_{ts}|=0.0395, |V_{ub}^{*}V_{us}|=0.0008.
\label{eq:shapewv}
\end{gather}
 We leave the CKM phase angle
$\gamma$ as a free parameter to explore  the branching ratio and
CP asymmetry parameter dependence on it.
 In SM, the CKM phase angle is the origin of CP violation.

\begin{figure}[htbp]
  \begin{center}
  \epsfig{file=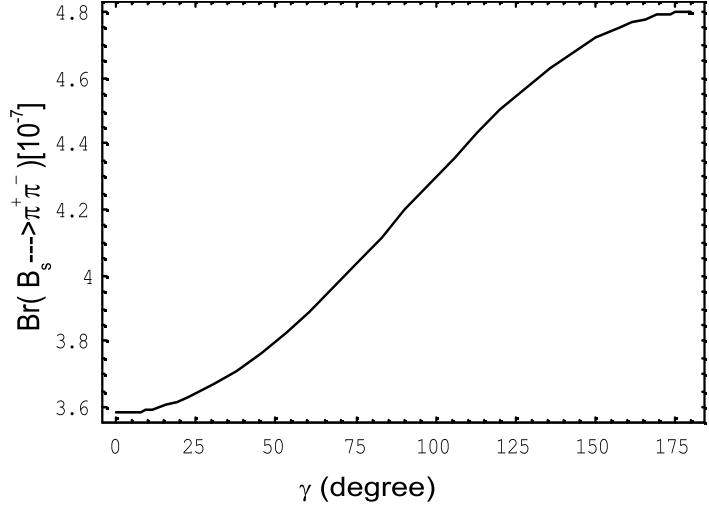,width=300pt,height=220pt}
   \end{center}
   \caption{The CP-averaged branching ratio  of
   $B_{s}^0(\bar{B}_{s}) \to \pi^{+} \pi^-$ decay as a function of CKM angle $\gamma$.}
   \label{figure:Fig2}
  \end{figure}

From Eq.(\ref{eq:width1}) and (\ref{eq:width2}), we get the averaged decay width for
$B_{s}^{0}(\bar{B}^{0}_{s})\rightarrow\pi^{+}\pi^{-}$
\begin{eqnarray}
 \Gamma(B_s^0(\bar{B}^{0}_{s}) \to \pi^+ \pi^-) &=& \frac{G_F^2 M_B^3}{128\pi}
(|A|^2/2+|\bar{A}|^2/2)\hspace*{1cm}  \nonumber \\
&=&\frac{G_F^2
M_B^3}{128\pi}|V_{ub}^*V_{us}M_{a}^{T}|^{2}[1+2z\cos\gamma\cos\delta+z^{2}].
\label{eq:width3}
\end{eqnarray}
Using the above parameters, we get $z=13.4$ and
$\delta=168^{\circ}$ in PQCD.
Eq.(\ref{eq:width3}) is a function of CKM angle $\gamma$. In Fig.
\ref{figure:Fig2}, we plot the averaged branching ratio of the
decay $B_{s}^{0}(\bar{B}^{0}_{s})\rightarrow\pi^{+}\pi^{-}$ with
respect to the parameter $\gamma$. From Fig. \ref{figure:Fig2}, we
can see that \footnote{Note that other  parameters such as
$m_{0\pi}$ and meson wave functions can  also
 lead to some uncertainties, their effects are already discussed in previous works \cite{8}. }:
\begin{equation}
\mathbf{Br} (B_s^0(\bar{B}^{0}_{s}) \to \pi^+ \pi^-)=(4.2\pm
0.6)\times10^{-7},
\end{equation}
for $0<\gamma<\pi$. The number
$z=|V_{tb}^*V_{ts}/V_{ub}^*V_{us}||M_{a}^{P}/M_{a}^{T}|=13.4$
means the amplitude of penguin diagrams is about 13.4 times more than that
of tree diagrams.
 Therefore almost all the contribution
comes from penguin diagrams in this decay and the branching ratio
is not sensitive to $\gamma$.

In Ref.\cite{17},  Beneke et al have estimated the branching ratio
for $B_s \to \pi^+\pi^{-}$ in QCD Factorization approach. In order
to avoid the endpoint singularities, they introduced parameters to
replace the divergent integral. In this approach, they estimated
that the branching ratio of this decay is
$(0.24-1.55)\times10^{-7}$ with those phenomenological parameters.
In our work, the calculation has no endpoint singularity because
of $k_T$\cite{6}. Our predicted result is larger than their simple
estimation, which can be tested by the experiments.

For the experimental side, we notice that there is only upper
limit of the decay $B_s^0\to\pi^+\pi^-$ given at $90\%$ confidence
level \cite{12}
\begin{equation}
\mathrm{Br}(B_s^0\to\pi^+\pi^-)< 1.4\times 10^{-4}.
\end{equation}
Obviously, our predicted result is still far from this upper limit.

\begin{figure}[htbp]
  \begin{center}
  \epsfig{file=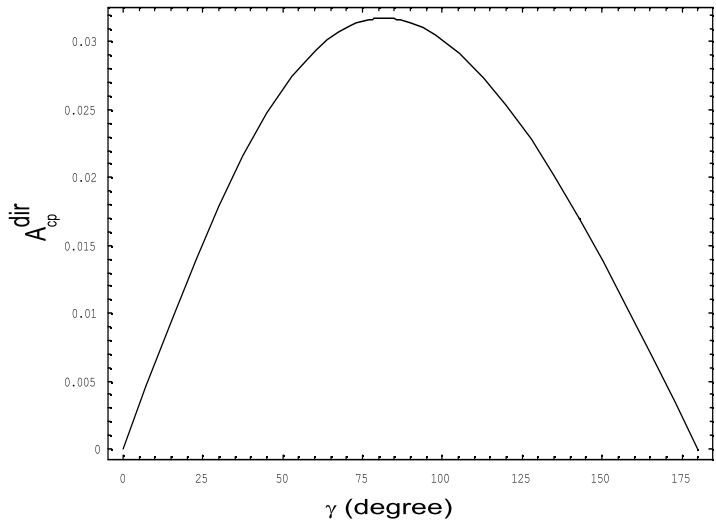,width=300pt,height=220pt}
   \end{center}
   \caption{Direct CP violation parameters of $B_{s}\to \pi^{+} \pi^-$ decay as a function of CKM angle $\gamma$.}
   \label{figure:Fig3}
  \end{figure}

  In SM, CP violation comes from interference between amplitudes
  with different CP eigenvalues. The strong interaction
  eigenstates $B_{s}^{0}$ and $\bar{B}_{s}^{0}$ can mix through
  weak interaction, i.e. $B_{s}^{0}$-$\bar{B}_{s}^{0}$ oscillation. By experimental observation we can know
  whether CP is  conserved. For the $B_{s}^{0}$-$\bar{B}_{s}^{0}$
  system, the CP asymmetry is time dependent \cite{3,13}:
  \begin{equation}
 A_{CP}(t)\simeq A_{CP}^{dir}\cos(\Delta
 m t)+a_{\epsilon+\epsilon^{'}}\sin(\Delta m t),\label{acp}
 \end{equation}
 where $\Delta m$ is the mass difference of the two mass
 eigenstates of $B_{s}$ mesons. $A_{CP}^{dir}$ is the direct CP
 violation parameter while $a_{\epsilon+\epsilon^{'}}$ is the
 mixing-related CP violation parameter. The direct CP violation
 parameter is defined as
 \begin{equation}
A_{CP}^{dir}=\frac{\Gamma(B_{s}^{0}\rightarrow\pi^{+}\pi^{-})-\Gamma(\bar{B}_{s}^{0}\rightarrow\pi^{+}\pi^{-})}{
\Gamma(B_{s}^{0}\rightarrow\pi^{+}\pi^{-})+\Gamma(\bar{B}_{s}^{0}\rightarrow\pi^{+}\pi^{-})}
=\frac{2z\sin\gamma\sin\delta}{1+2z\cos\gamma\cos\delta+z^{2}} .
\label{dcpv}
 \end{equation}
 Using Eq.(\ref{eq:width1}) and (\ref{eq:width2}), we can compute the parameter
 $A_{CP}^{dir}$. The direct CP asymmetry  $A_{CP}^{dir}$ has a strong dependence
 on the CKM angle, as can be seen easily from Eq.(\ref{dcpv}) and Fig. \ref{figure:Fig3}.
 From this figure one can see that
 when the CKM angle $\gamma$ is around $\pi /2$ the direct CP asymmetry
 reaches its peak, which is about $3\%$. The small direct CP
 asymmetry is also a result of small tree level contribution.

\begin{figure}[htbp]
  \begin{center}
  \epsfig{file=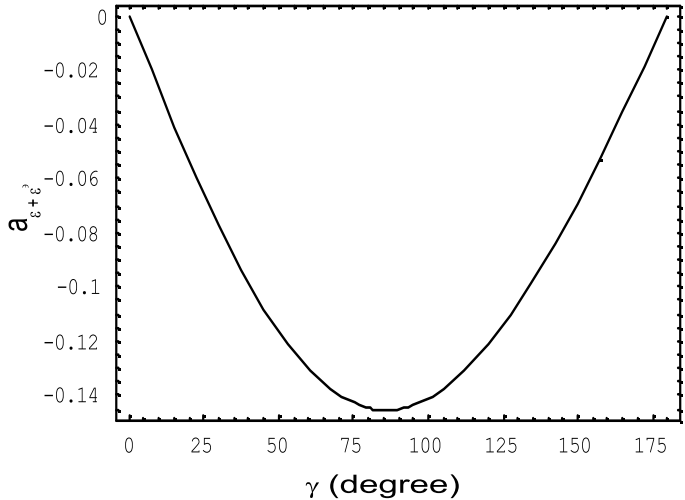,width=300pt,height=220pt}
   \end{center}
   \caption{Mixing-related CP violation parameter $a_{\epsilon+\epsilon^{'}}$
    of $B_{s}^0(\bar{B}_{s}) \to \pi^{+} \pi^-$ decay as a function of CKM angle $\gamma$.}
   \label{figure:Fig4}
  \end{figure}

  The mixing-related CP violation parameter in Eq.(\ref{acp}) is defined
  as \cite{9}
 \begin{equation}
a_{\epsilon+\epsilon^{'}}=\frac{-2Im(\lambda_{CP})
}{1+|\lambda_{CP}|^{2}},
 \end{equation}
 where
 \begin{equation}
\lambda_{CP}=\frac{V_{tb}^{*}V_{ts}\langle
f|H_{eff}|\bar{B}_{s}^{0}\rangle}{V_{tb}V_{ts}^{*}\langle
f|H_{eff}|B_{s}^{0}\rangle}          .
 \end{equation}
 In Fig. \ref{figure:Fig4}, we study the mixing CP violation parameter $a_{\epsilon+\epsilon^{'}}$  of the
 decay $B_{s}\rightarrow\pi^{+}\pi^{-}$  as a function of CKM angle
 $\gamma$, just like the case of direct CP violation, it is almost
 symmetric and the symmetry axis is near $\gamma=\pi/2$. its peak
 is close to $-14.5\%$. The possible large CP asymmetry might be observed at
 LHCb experiment in the future, this would help us to determine the
 value of CKM angle $\gamma$.

  In conclusion, we study the branching ratio and CP asymmetry of
  the decay $B_{s}^{0}(\bar{B}^{0}_{s})\rightarrow\pi^{+}\pi^{-}$
  in PQCD, we find that the branching ratio is at the
  order of $10^{-7}$ and there are  large CP asymmetries in the
  process, which may be measured in the future LHC-b experiments
  and BTeV experiment at Fermilab.
  This small branching ratio, predicted in the SM, make it
  sensitive to possible new physics contribution.

\section*{Acknowledgments}

 This work is partly supported by National Science Foundation of
 China under Grant (No. 90103013, 10135060 and 10275035). Y.Li thanks M-Z Yang for useful discussions.


\begin{appendix}

\section{Some formulas used in the text}

For $B_{s}$
meson wave function, we use the same wave function as
  $B$ meson \cite{9,14}, despite the possible SU(3) breaking
  effect
\begin{equation}
\phi_{B_s}(x,b) = N_{B_s} x^2(1-x)^2 \exp \left[ -\frac{M_{B_s}^2\
x^2}{2 \omega_b^2} -\frac{1}{2} (\omega_b b)^2
\right].\label{waveb}
\end{equation}
The parameter  $\omega_b=0.4\mbox{ GeV}$,
is constrained by other charmless
B decays \cite{9,14}.                          And
$N_{B_s}=114.0\mbox{GeV}$  is the normalization constant
using $f_{B_s}=236 \mbox{MeV}$.

The $\pi$ meson's distribution amplitudes are given by light cone
QCD sum rules \cite{15}:
\begin{eqnarray}
\nonumber \phi_{\pi}^A(x) &=& \frac{3f_{\pi}}{\sqrt{2 N_c}} x(1-x)
\left\{ 1 +0.44C_2^{3/2}(t) + 0.25C_4^{3/2}(t) \right\}, \\
\nonumber \phi_{\pi}^P(x) &=& \frac{f_{\pi}}{2\sqrt{2 N_c}}
\left\{ 1 +0.43C_2^{1/2}(t) +
0.09C_4^{1/2}(t) \right\}, \\
\phi_{\pi}^T(x) &=& \frac{f_{\pi}}{2\sqrt{2 N_c}}(1-2x) \left\{
1+0.55(10x^2-10x+1) \right\},
\end{eqnarray}
where $t=1-2x$. The Gegenbauer polynomials are defined by:
\begin{eqnarray}
C_2^{1/2}(t)=\frac{1}{2}(3t^2-1),\hspace{1cm} C_4^{1/2}(t)=\frac{1}{8}(35t^4-30t^2+3),\nonumber\\
C_2^{3/2}(t)=\frac{3}{2}(5t^2-1),\hspace{1cm}
C_4^{3/2}(t)=\frac{15}{8}(21t^4-14t^2+1).
\end{eqnarray}

Since the hard part is calculated only to leading order of $\alpha_s$,
we use the one loop expression for the strong running coupling
constant in our numerical analysis,
\begin{equation}
 \alpha_s(\mu)=\frac{4\pi}{\beta_0\mathrm{ln}(\mu^2/\Lambda^2)},
\end{equation}
where $\beta_0=(33-2n_{f})/3$ and $n_{f}$ is the number of active
quark flavor at the appropriate scale $\mu$. $\Lambda$ is the QCD
scale, we set $\Lambda=250 $MeV at $n_{f}=4$.

The function $E_{m}(t_{i})$ in  Eq.(\ref{ma}) are
defined by
\begin{gather}
E_{m}(t_{i}) = e^{-S_B(t_{i})-S_{\pi^{+}}(t_{i})-S_{\pi^{-}}(t_{i})}
,
\end{gather}
where $S_{B}$, $S_{\pi^{+}}$, $S_{\pi^{-}}$ result from
summing both double logarithms due to  infrared gluon corrections and
single ones caused by the renormalization of ultra-violet
divergence. They are defined as:
\begin{gather}
 S_B(t)=s(x_1P_1^+,b_1)+2\int_{1/b_1}^t\!\!\!\frac{d\bar\mu}{\bar\mu}\gamma(\alpha_s(\bar\mu)),\hspace*{3.5cm}\\
 S_{\pi^-}(t)=s(x_2P_2^+,b_2)+s((1-x_2)P_2^+,b_2)+
 2\int_{1/b_2}^t\!\!\!\frac{d\bar\mu}{\bar\mu}\gamma(\alpha_s(\bar\mu))\\
 S_{\pi^+}(t)=s(x_3P_3^-,b_3)+s((1-x_3)P_3^-,b_3)+
 2\int_{1/b_3}^t\!\!\!\frac{d\bar\mu}{\bar\mu}\gamma(\alpha_s(\bar\mu)),
\end{gather}
where $s(Q,b)$ called Sudakov factor is given as \cite{16}
\begin{equation}
s(Q,b)=\int_{1/b}^Q\!\!\! \frac{d\mu}{\mu}\Bigl[
\ln\left(\frac{Q}{\mu}\right)A(\alpha(\bar\mu))+B(\alpha_s(\bar\mu)) \Bigr]
\label{su1}
\end{equation}
with
\begin{gather}
A=C_F\frac{\alpha_s}{\pi}+\left[\frac{67}{9}-\frac{\pi^2}{3}-\frac{10}{27}n_{f}+
\frac{2}{3}\beta_0\ln\left(\frac{e^{\gamma_E}}{2}\right)\right]
 \left(\frac{\alpha_s}{\pi}\right)^2 ,\\
B=\frac{2}{3}\frac{\alpha_s}{\pi}\ln\left(\frac{e^{2\gamma_{E}-1}}{2}\right)\hspace{6cm}
\end{gather}
here $\gamma_E$ is the Euler constant, $n_{f}$ is the active
flavor number.

The functions $h_a^{(1)}$, and $h_a^{(2)}$  in  Eq.(\ref{ma}) come from
the Fourier transformation of propagators of virtual quark and gluon. They are defined by
\begin{align}
& h^{(j)}_a(x_1,x_2,x_3,b_1,b_2) = \nonumber \\
& \biggl\{ \frac{\pi i}{2} \mathrm{H}_0^{(1)}(M_B\sqrt{x_2x_3}\,
b_1)
 \mathrm{J}_0(M_B\sqrt{x_2x_3}\, b_2) \theta(b_1-b_2)
\nonumber \\
& \qquad\qquad\qquad\qquad + (b_1 \leftrightarrow b_2) \biggr\}
 \times\left(
\begin{matrix}
 \mathrm{K}_0(M_B F_{(j)} b_1), & \text{for}\quad F^2_{(j)}>0 \\
 \frac{\pi i}{2} \mathrm{H}_0^{(1)}(M_B\sqrt{|F^2_{(j)}|}\ b_1), &
 \text{for}\quad F^2_{(j)}<0
\end{matrix}\right),
\label{eq:propagator2}
\end{align}
where $\mathrm{H}_0^{(1)}(z) = \mathrm{J}_0(z) + i\,
\mathrm{Y}_0(z)$, and $F_{(j)}$s are defined by
\begin{equation}
 F^2_{(1)} = x_1x_3-x_2x_3,\
F^2_{(2)} = x_1 +x_2+x_3-x_1x_3-x_2x_3.
\end{equation}

The hard scale $t_{i}(i=1,2)$ in Eq.(\ref{ma}) are taken as the largest
energy scale in the $H$ to kill the large logarithmic radiative
corrections:
\begin{gather}
 t_i = \mathrm{max}(M_B \sqrt{|F^2_{(i)}|},
M_B \sqrt{x_2x_3 }, 1/b_1,1/b_2)\hspace{1cm} (i=1,2).
\end{gather}

\end{appendix}


\end{document}